\documentclass[prl,amssymb,nobibnotes,aps,superscriptaddress,floatfix,lengthcheck,showpacs,longbibliography,nofootinbib]{revtex4-1}
\usepackage[utf8]{inputenc}
\usepackage[T1]{fontenc}
\usepackage{amsmath}
\usepackage{enumitem}
\usepackage{euscript}
\usepackage{graphicx}
\usepackage{amsfonts}
\usepackage{subfigure}
\usepackage[usenames]{color}

\usepackage{microtype}
\usepackage[colorlinks,linkcolor=blue,citecolor=blue,urlcolor=blue]{hyperref}

\def\equationautorefname~#1\null{Eq.~(#1)\null}

\begin{document}
\title{Imaging ultrafast molecular wavepackets with a single chirped UV pulse}
\author{Denis Jelovina}
\affiliation{Departamento de Qu\'imica, Universidad Aut\'onoma de Madrid, 28049 Madrid, Spain}
\author{Johannes Feist}
\affiliation{Departamento de F\'isica Te\'orica de la Materia Condensada, Universidad Aut\'onoma de Madrid, 28049 Madrid, Spain}
\affiliation{Condensed Matter Physics Center (IFIMAC), Universidad Aut\'onoma de Madrid, 28049 Madrid, Spain}
\author{Fernando Mart\'in}
\affiliation{Departamento de Qu\'imica, Universidad Aut\'onoma de Madrid, 28049 Madrid, Spain}
\affiliation{Condensed Matter Physics Center (IFIMAC), Universidad Aut\'onoma de Madrid, 28049 Madrid, Spain}
\affiliation{Instituto Madrile\~no de Estudios Avanzados (IMDEA) en Nanociencia, 28049 Madrid, Spain}
\author{Alicia Palacios}
\email{alicia.palacios@uam.es}
\affiliation{Departamento de Qu\'imica, Universidad Aut\'onoma de Madrid, 28049 Madrid, Spain}
\date{\today}
\begin{abstract}
We show how to emulate a conventional pump-probe scheme using a single frequency-chirped ultrashort UV pulse to obtain a time-resolved image of molecular ultrafast dynamics. The chirp introduces a spectral phase in time that encodes the delay between the pump and the probe frequencies contained in the pulse. By comparing the results of full dimensional ab initio calculations for the H$_2^+$ molecule with those of a simple sequential model, we demonstrate that, by tuning the chirp parameter, two-photon energy-differential ionization probabilities directly map the wave packet dynamics generated in the molecule. As a result, one can also achieve a significant amount of control of the total ionization yields, with a possible enhancement by more than an order of magnitude.
\end{abstract}
\maketitle
The advent of free-electron-laser facilities and high-harmonic generation has opened the way to the production of intense and ultrashort ultraviolet (UV) pulses with durations in the femtosecond and attosecond range \cite{Kling2008,Krausz2009,Bostedt2009,Bostedt2016}. One of the more awaited capabilities offered by such pulses is to use them to monitor and control electronic and nuclear dynamics, for example within a UV-UV pump-probe scheme, the so-called ``holy grail'' of attosecond physics. While some progress in this direction has been made~\cite{Tzallas2011,Wostmann2013,Carpeggiani2014,Campi2016,Takanashi2017}, there are many technical challenges still to overcome, such as the limited intensity of the pulses and the fact that few optically active elements exist in the (extreme) ultraviolet. This precludes, for example, the use of pulse-shaping techniques and coherent control approaches that can be applied at optical and infrared frequencies to produce an ``optimal'' pulse for a desired photo-induced physical process or chemical reaction \cite{Assion1998, Meshulach1998, Weiner2000, Levis2001, Brixner2004, Chatel2005, Djotyan2004, Weiner2000, Nuernberger2009}. Consequently, most experiments performed so far with attosecond UV pulses rely instead on an intense infrared pulse for either the pump or the probe step, which can significantly distort the system and alter the dynamics.

In this Letter, we demonstrate that by changing a single parameter, the spectral chirp of an ultrashort UV pulse, we can achieve a significant amount of control over molecular multiphoton ionization, changing the total ionization yield by more than a factor of ten. More importantly, we show how to emulate a conventional pump-probe setup to obtain direct time-resolved imaging of ultrafast molecular dynamics. The spectral chirp is experimentally tuneable both in high harmonic generation and with free electron lasers \cite{Chang2005,Scrinzi2006,Hofstetter2011,Bostedt2016}. A similar idea was previously suggested by Yudin et al.~\cite{Yudin2006}, but only demonstrated for a superposition of two bound states in the hydrogen atom. In the present work, we aim at reconstructing the vibronic wave packet in a small molecule, simultaneously pumped and probed by a single chirped UV pulse.

At optical and infrared frequencies, the effect of frequency-chirped pulses has been actively investigated using theoretical approaches based on second-order time-dependent perturbation theory (TDPT) to treat few-photon excitation processes in atoms \cite{Adler1995,Zhang2009,Felinto2009,Dudovich2001,Meshulach1998,Meshulach1999,Chatel2003,Djotyan2004,Stowe2006}. These methods have also been applied at ultraviolet frequencies, but, to our knowledge, only to describe the chirp-dependent photoelectron angular distributions in atomic photoionization \cite{Pronin2009,Peng2009,Pronin2011}. In these approaches, only a limited number of states partake in the dynamics. In contrast, the additional nuclear degrees of freedom in molecular targets, as treated here, induce more complex wavepacket motion characterized by the participation of many vibronic states. We thus directly solve the full-dimensional time-dependent Schr\"odinger equation (TDSE), $-i\frac{\partial}{\partial t} \Phi(t) = H(t) \Phi(t)$, using H$_2^+$ as a benchmark target to investigate the coherent manipulation of two-photon molecular photoionization by chirped pulses.

In our approach, the time-dependent wave function $\Phi(t)$ is expressed within a single-center expansion, using spherical harmonics to treat the angular components and a finite element discrete variable representation (FEDVR) for the radial coordinates of \emph{both} the electronic and vibrational degrees of freedom. The full Hamiltonian, $H(t) = H_0 + V(t)$, is given by the sum of the Hamiltonian of the isolated molecule, $H_0$, which depends on both electronic and nuclear coordinates and therefore implicitly incorporates non-adiabatic couplings, and the laser-molecule interaction within the dipole approximation in length gauge, $\mathbf{V}(t)=\mathbf{r}\cdot\mathbf{E}(t)$, as the product of the electronic coordinates $\mathbf{r}$ and the electromagnetic field of the pulse $\mathbf{E}(t)$.  After solving the TDSE, we implicitly propagate until infinite time and simultaneously Fourier-transform the time-dependent wave function to obtain the scattering function at a given energy \cite{Palacios2007}. From the scattering wave function, one can extract the single ionization amplitudes by using the surface integral formalism described in \cite{Palacios2007} and thus obtain the total, energy- and angle-differential ionization probabilities.  The electromagnetic field $\mathbf{E}(t)$ of a (linearly polarized) chirped Gaussian pulse can be written as \cite{Yudin2006,Nakajima2007,Peng2009}:
\begin{equation}
  \mathbf{E}(t) = \frac{1}{2} E_{\mathrm{max}}(\eta) F(t) \exp(i \phi(\eta,t)) \mathbf{e}_z + c.c.,
  \label{eq:chirped_gaussian}
\end{equation}
where the instantaneous phase is given by
\begin{equation}
  \phi(\eta,t) = \omega_0 t - \frac{\eta}{2 T_0^2 (1+\eta^2)} t^2,
\end{equation}
and the temporal envelope $F(t)$ is described by a Gaussian function:
\begin{equation}
  F(t) = \exp\left(-\frac{t^2}{2T(\eta)^2}\right) .
\end{equation}
The field amplitude, $E_{\mathrm{max}}(\eta)$, the pulse duration $T(\eta)$ and the instantaneous frequency $\omega(\eta,t) = \frac{\partial}{\partial t} \phi(\eta,t)$ explicitly depend on the chirp parameter $\eta$.
Note that the spectral chirp (the quadratic term of the spectral phase) of the field defined here is directly proportional to $\eta$, but the temporal chirp [prefactor of the $t^2$ term in $\phi(\eta,t)$] has opposite sign in contrast with the definition in \cite{Yudin2006,Nakajima2007,Peng2009}.
For unchirped pulses ($\eta=0$), $E_{max}(\eta=0)=E_0$ is the peak
amplitude, $T_0$ defines the duration of the pulse (FWHM of the field
envelope is $T_{\mathrm{FWHM}}=2\sqrt{\log 4}\,T_0$), and $\omega_0$
is the carrier frequency. The parametrization is chosen such that adding a chirp in frequency ($\eta\neq 0$), the spectrum remains unchanged.
The duration of the pulse
then increases to $T(\eta) = T_0 \sqrt{1+\eta^2}$, while the peak amplitude
decreases to $E_{\mathrm{max}}(\eta) = E_0 / (1+\eta^2)^{1/4}$.
In other words, the same frequencies are ``stretched'' over a longer duration. The Fourier transform of this Gaussian pulse leads to a Gaussian spectral function, with amplitude independent of $\eta$, but a spectral phase that is quadratic in $\omega$ with a spectral chirp given by $\eta T_0^2/2$.

\begin{figure}[tb]
\begin{center}
\includegraphics[width=1.0\columnwidth, clip=true,angle=00]{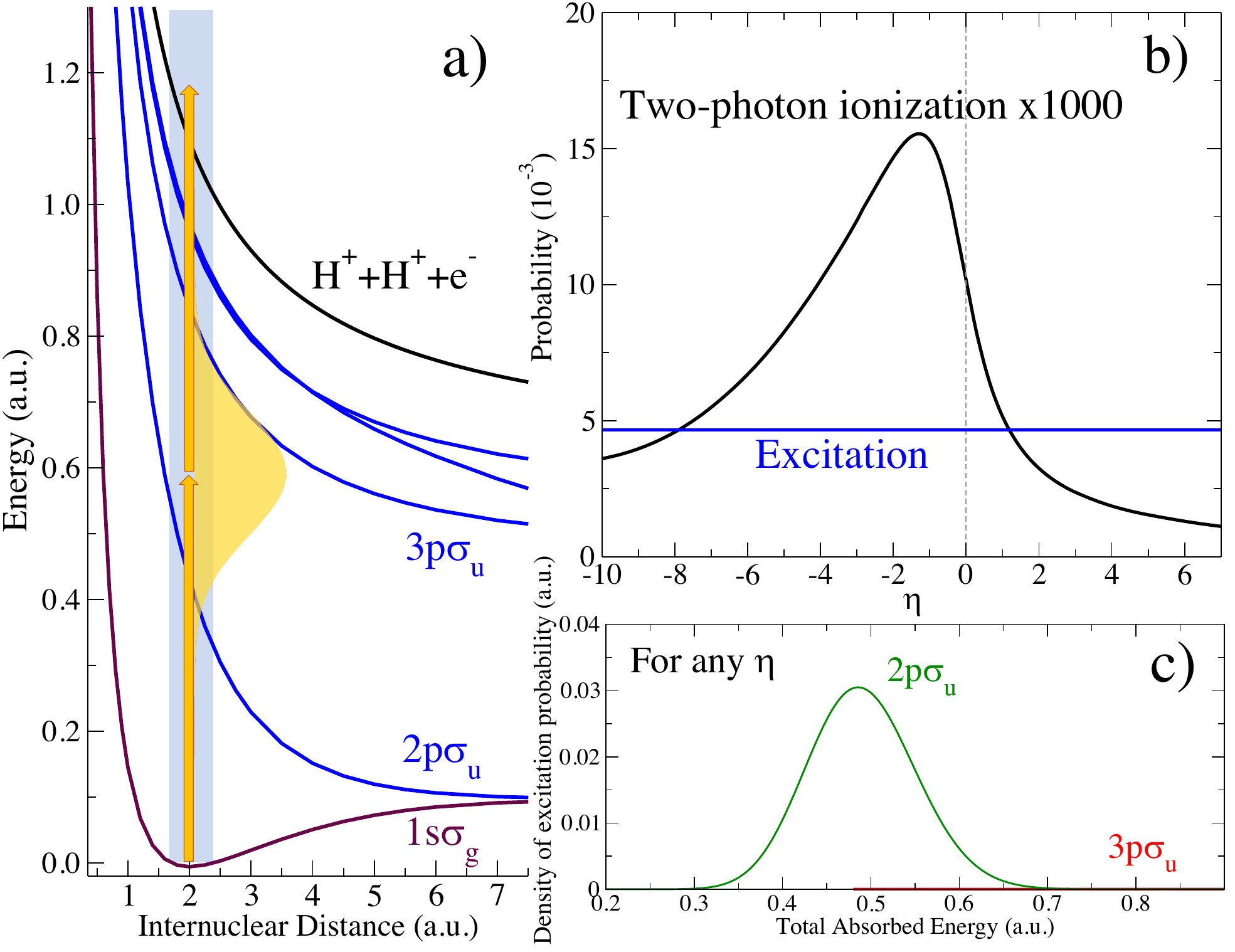}
\caption{(a) Energy scheme with the relevant potential energy curves: ground state of H$_2^+$(1s$\sigma_g$) in violet, first four excited states of $\sigma_u$ symmetry in blue and the Coulomb explosion potential in black. The energy bandwidth of the pulses employed in the present work is plotted in an orange shadowed area in the region where the one-photon absorption occurs, centered at 0.6 a.u.\ and covering an energy range around 0.4-0.8 a.u.\ The blue shadowed area indicates the Franck-Condon region. (b) Two-photon ionization (black) and one-photon excitation (blue) yields as a function of the chirp parameter $\eta$. (c) One-photon excitation distributions as a function of the total absorbed energy for the two lowest excited states 2p$\sigma_u$ (green) and 3p$\sigma_u$ (red). }
\label{fig:fig1}
\end{center}
\end{figure}

In \autoref{fig:fig1}(a) we show the energetics of the two-photon ionization process using linearly polarized light parallel to the molecular axis of the H$_2^+$ molecule. The one-photon transition from the 1s$\sigma_g$ ground state creates a vibronic wave packet in the dissociative excited states of $\sigma_u$ symmetry. The two-photon transition reaches the ionization potential leading to the Coulomb explosion of the system (H$^+$+H$^+$+e$^-$).
We use pulses whose frequency spectrum corresponds to that of an unchirped pulse ($\eta=0$) with a FWHM duration of $450$\ as centered at $\omega_0=0.6$~a.u.\ and a laser intensity of $1.1 \times 10^{13}$\ W/cm$^2$. With these parameters, ionization is solely due to two-photon absorption paths. The energy bandwidth of these pulses is plotted in \autoref{fig:fig1}(a) as an orange shadowed area. The maximum spectral amplitude, 0.6 a.u., lies in between the 2p$\sigma_u$ and 3p$\sigma_u$ states.

As expected, the excitation probability, shown in \autoref{fig:fig1}(b), is independent of the chirp parameter. As graphically described by Brumer and Shapiro \cite{Brumer1989}, the one-photon absorption probability is ``\emph{an emperor without clothes}'', unaffected by the spectral phase, and only depends on the spectral frequency distribution of the pulse. The spectral phase introduced in the excited wave packet can only be captured in a second-order process, for instance the two-photon transition depicted in \autoref{fig:fig1}, where the time evolution of the nuclear wave packet is retrieved through its projection into the electronic continua.
The excitation probabilities associated to the 2p$\sigma_u$ and 3p$\sigma_u$ states, which are independent of $\eta$, are plotted as a function of the vibronic (vibrational+electronic) energy in \autoref{fig:fig1}(c). For the pulses employed here, we can see that the two-photon ionization proceeds almost entirely through the first excited state. First, the photon energies within the pulse are energetically closer to the resonant vertical transition from the ground state to the 2p$\sigma_u$ state. In addition, the dipole coupling to the 3p$\sigma_u$ state is noticeably weaker than that to 2p$\sigma_u$. As a result, the one-photon excitation probability to the 3p$\sigma_u$ state is three orders of magnitude smaller than that to the 2p$\sigma_u$ state.

The total ionization probability as a function of the $\eta$ parameter is also included in \autoref{fig:fig1}(b). As mentioned above, ionization to the final states of $\Sigma_g$ symmetry (even number of absorbed photons) is the dominant process, while the ionization to states of $\Sigma_u$ symmetry (odd number of absorbed photons) is negligible.
As shown in the figure, by tuning the chirp parameter the total ionization probability can be strongly modified, with a modulation range of more than an order of magnitude.
At the H$_2^+$ equilibrium distance, the energy difference between the ground and the 2p$\sigma_u$ state is $0.43$~a.u., while the difference between the latter and the Coulomb explosion potential energy curve is $0.67$~a.u.. It is thus expected that the total ionization yield is enhanced for negative values of $\eta$, i.e. when lower frequencies ($<\!0.6$~a.u.) arrive earlier and larger frequencies ($>\!0.6$~a.u.) arrive later. In this way, both transitions, from the ground state to 2p$\sigma_u$ and from 2p$\sigma_u$ to the Coulomb explosion, can take place when the instantaneous frequency is close to resonant, thus maximizing ionization.
\begin{figure}[tb]
\begin{center}
\includegraphics[width=1.0\columnwidth, clip=true,angle=00]{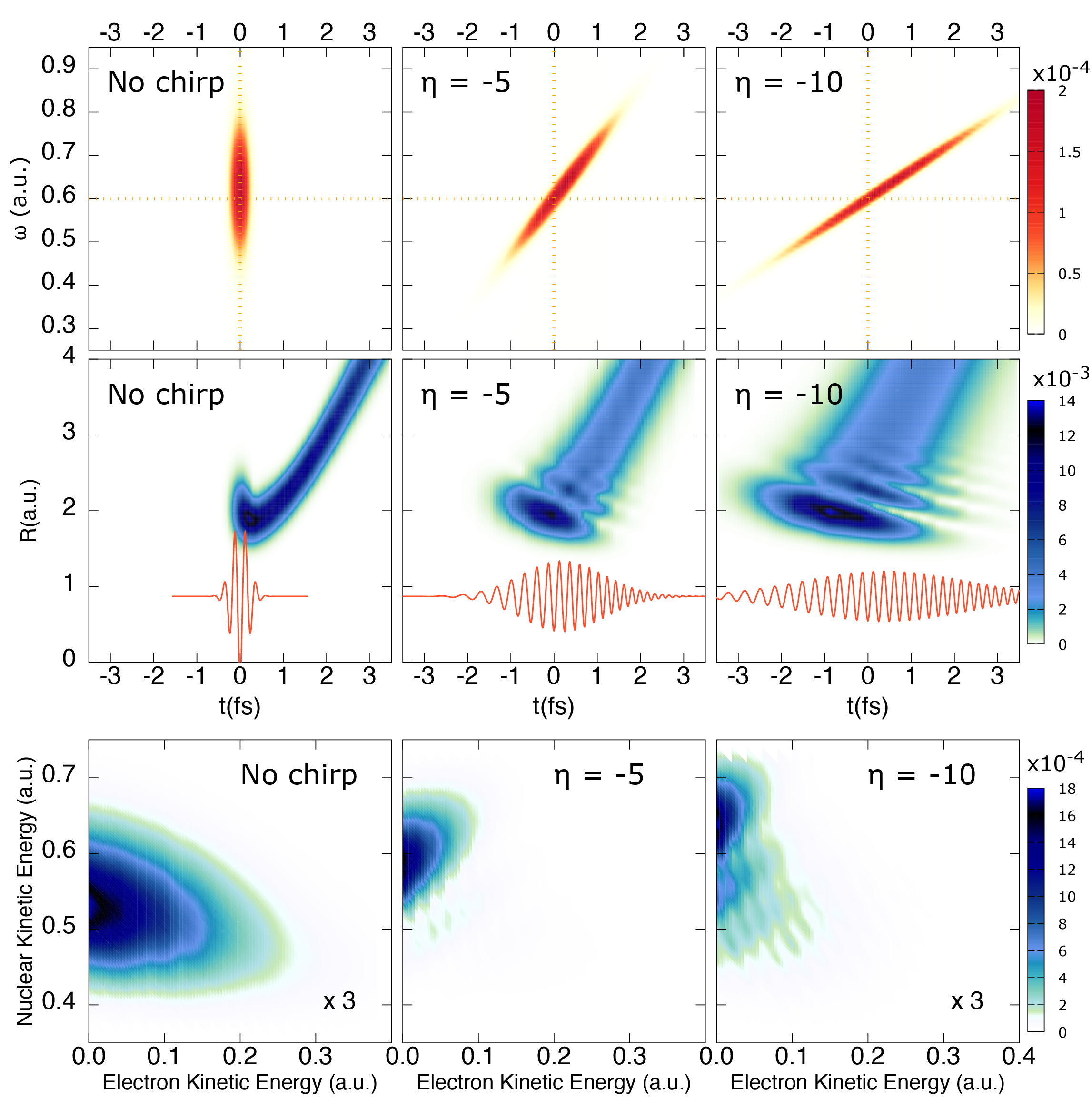}
\caption{Results for three different values of the chirp parameter ($\eta=0$, $-5$, and $-10$ as labeled in each subplot). Upper row: Wigner distributions. Middle row: Nuclear wave packet associated to the 2p$\sigma_u$ excited state as a function of time. The electromagnetic field of the pulse, $E(t)$, is included for each $\eta$ (red line). Lower row: Fully differential energy distributions for the ionized fragments after Coulomb explosion ($x$-axis: electronic energy, $y$-axis: nuclear energy).}
\label{fig:fig2}
\end{center}
\end{figure}

In order to extract dynamical information about the excited wave packet associated to the 2p$\sigma_u$ state, we will study the energy-differential ionization probabilities for different values of the chirp parameter. In the upper row of \autoref{fig:fig2}, we plot the Wigner distributions of the electromagnetic field, which provides a combined time-frequency representation, for three different pulses with $\eta=0$, $-5$ and $-10$. For the unchirped pulse, all frequencies reach the target simultaneously. However, for the chirped pulses, the more negative $\eta$ the larger the time delay between the lower and the higher frequencies.
In other words, by making $\eta$ more negative, we are creating a nuclear wave packet in the 2p$\sigma_u$ state at earlier times (the direct vertical transition at $0.43$~a.u.\ occurs earlier), which is probed by promotion into the Coulomb explosion channel at later times (frequencies around $0.67$~a.u.\ arrive later).
Therefore, this is conceptually equivalent to standard pump-probe schemes, where two time-delayed pulses are employed: one pulse launches the dynamics in the target and a second pulse, delayed (and ideally not overlapping) in time, probes the pumped dynamics through promotion to a given final state. In the present case, the time delay is encoded in the chirp parameter.

The middle panels of \autoref{fig:fig2} show the corresponding nuclear wave packets (NWPs) in the 2p$\sigma_u$ state as a function of time ($x$-axis) and internuclear distance ($y$-axis). In the same subplots, we include the electromagnetic field, $E(t)$, as a red line. We can see that the quadratic spectral phase associated to a given chirp value ($\eta\ne0$) introduces structure in the pumped excited wave packet. This is due to interferences resulting from different frequency components with different spectral phases \cite{Chatel2005}.
We observe nearly the same wave packet, but stretched in time. As discussed above (cf.~\autoref{fig:fig1}c), the energy distribution of the wave packet is identical for all values of $\eta$. However, as seen in \autoref{fig:fig2}, their spatial structure differs, since
for the more negative chirp, the same frequencies are reaching the target with a larger delay between them.
This structured wave packet is mapped into the energy differential ionization probabilities upon absorption of a second photon, leading to distinct profiles.
\begin{figure}[tb]
\begin{center}
\includegraphics[width=1.0\columnwidth, clip=true,angle=00]{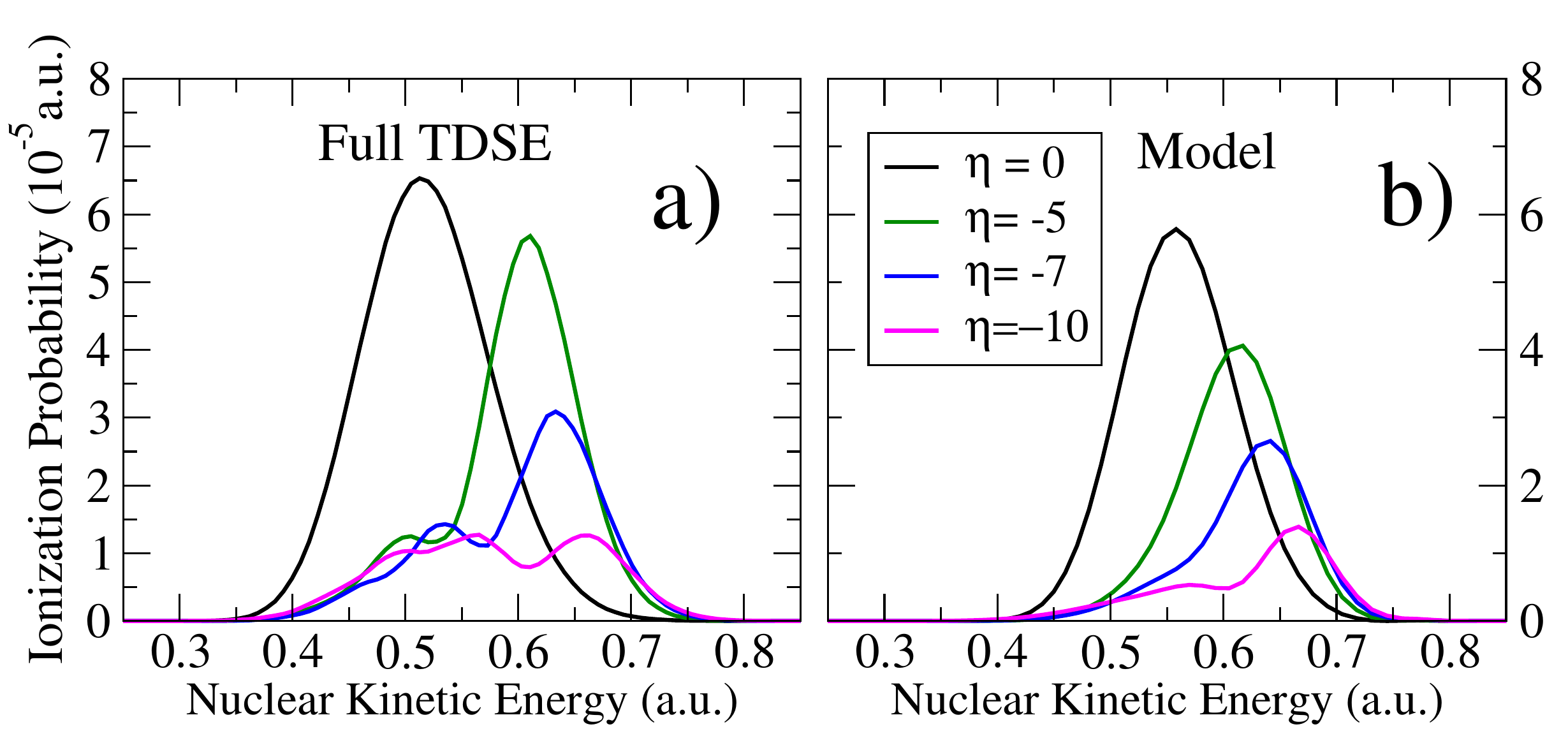}
\caption{Ionization probability as a function of the nuclear kinetic energy release for different values of the chirped parameters (see legend), extracted from the full dimensional calculation solving the TDSE (a) and extracted from the ``sequential'' model based on second order time-dependent perturbation theory as explained in the text (b).}
\label{fig:fig3}
\end{center}
\end{figure}
The energy-differential ionization probabilities are shown in the contour plots in the bottom panels of \autoref{fig:fig2}, as a function of the ejected electron energy ($x$-axis) and the nuclear kinetic energy release of the nuclei ($y$-axis). The energy distribution resulting from the interaction with the unchirped pulse is smooth, while the chirped pulses yield distributions shifted towards higher nuclear kinetic energies and with internal structure. For a better visualization, we integrate the ionization probabilities over the electron kinetic energy and obtain the nuclear kinetic energy distributions shown in \autoref{fig:fig3}(a). Here, we have additionally included the results for $\eta=-7$. These energy distributions actually reflect the dynamics launched in the excited molecule. In order to prove this,
in \autoref{fig:fig3}(b), we show the results of a sequential model where the excited NWP created in the 2p$\sigma_u$ state by the lower frequencies is directly projected into the ionization channel. Note that the model qualitatively reproduces the position and the profile of each energy distribution.
The model uses as starting point the exact second-order time-dependent perturbation theory expression for the molecular wave packet, $\Psi^{(2)}_I(t)$, created after two-photon absorption from the ground state, $\Psi_0$. In the interaction picture, it is given by
\begin{equation}
| \Psi^{(2)}_I(t) \rangle = \frac1i \int_{-\infty}^{t} dt' \hat{V}_I(t') |\Psi^{(1)}_I(t') \rangle,
\label{eq:2TDPT}
\end{equation}
\begin{equation}
| \Psi^{(1)}_I(t') \rangle = \frac1i \int_{-\infty}^{t'} dt'' \hat{V}_I(t'') |\Psi_0 \rangle,
\label{eq:1TDPT}
\end{equation}
where $\hat{V}_I(t)=e^{iH_0t}V(t)e^{-iH_0t}$ is the driving operator in the interaction picture. The ionization amplitude can be obtained by simply projecting the molecular wave packet, $|\Psi^{(2)}_I(t)\rangle$, into the final continuum states, leading to the ionization probabilities in the bottom panel of \autoref{fig:fig2}. The first-order wavepacket $|\Psi^{(1)}_I(t') \rangle$ corresponds to the nuclear wavepacket after one-photon absorption, as shown in the middle panels of \autoref{fig:fig2} (note that those NWPs are plotted in the Schr\"odinger picture and consequently evolve in time even in the absence of the field). In the interaction picture, the wavepackets remain unchanged in time once the frequency components of the driving pulse that are responsible for the transition have been absorbed. The ab initio first-order wavepackets at $t\to\infty$ are shown in \autoref{fig:fig4}(a). For negative chirps, these wavepackets are already fully formed when the second (higher-frequency) photon is absorbed. We can thus use a sequential approximation where the final first-order wavepacket (with $t\to\infty$) is used as the source for the second-order wavepacket:
\begin{equation}
| \Psi^{(2)}_\mathrm{seq}(t)\rangle= \frac1i \int_{-\infty}^{t} dt' \hat{V}_I(t') | \Psi^{(1)}_I(t'\to\infty) \rangle .
\label{eq:2TDPT_seq}
\end{equation}
\begin{figure}[tb]
\begin{center}
\includegraphics[width=1.0\columnwidth, clip=true,angle=00]{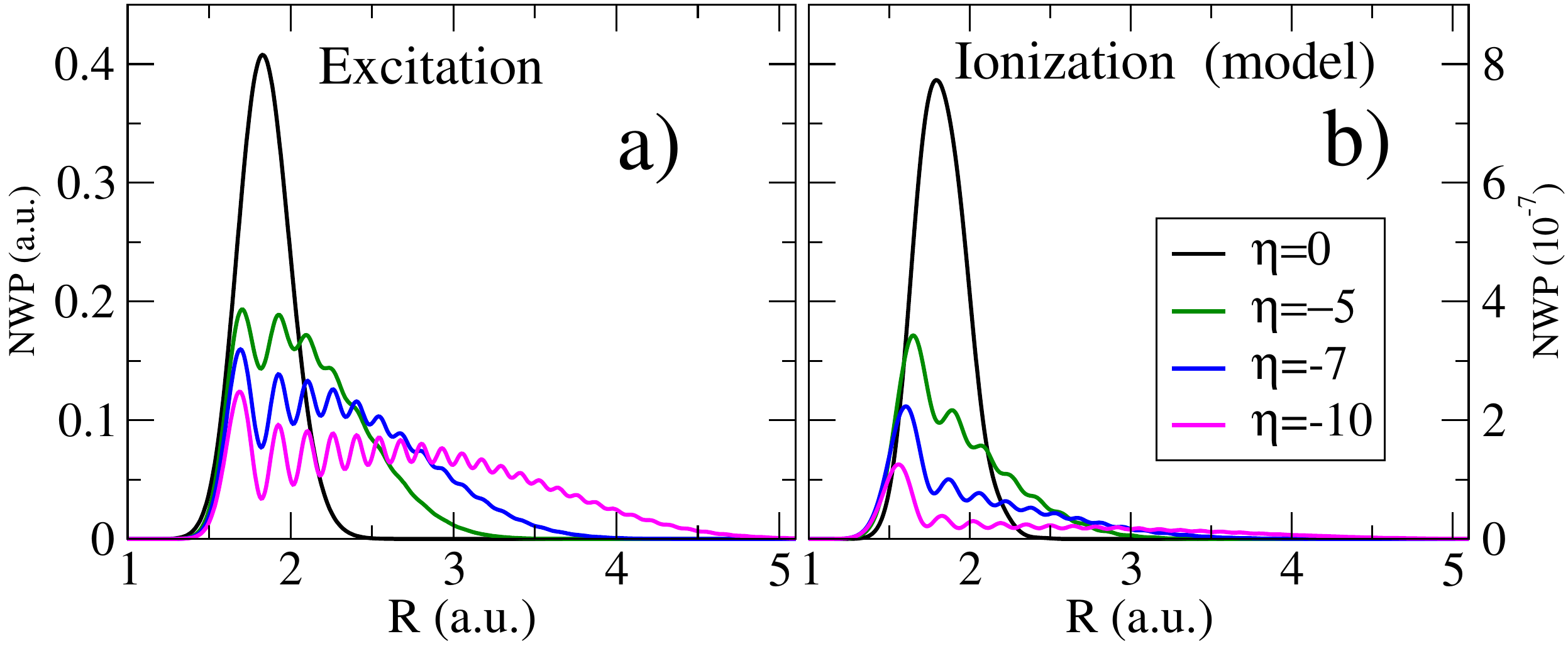}
\caption{Nuclear wave packets as a function of internuclear distance for different values of $\eta$ indicated in the legend. (a) Ab initio calculated excitation nuclear wave packet in the interaction picture at the end of the pulse. (b) Mapping of the excitation wave packet plotted on the left into the Coulomb explosion potential energy curve using the model explained in the text.}
\label{fig:fig4}
\end{center}
\end{figure}

The result of this approximation is plotted in \autoref{fig:fig4}(b), where we can see how the structure of the excited wave packet in \autoref{fig:fig4}(a) is reflected in the ionized wave packet extracted from the model. By using the definition of $\hat{V}_I(t)$ and \autoref{eq:chirped_gaussian}, the corresponding ionization amplitude, i.e. the projection of the approximated second-order wave packet into the final states, $c_f = \langle f | \Psi^{(2)}_\mathrm{seq}(t\to\infty)\rangle$, can be written as:
\begin{equation}
c_f \propto \sum_n a_{fn}a_{ni} e^{-i\frac{\eta T_0^2}{2}\left[(\omega_{fn}-\omega_0)^2+(\omega_{ni}-\omega_0)^2\right]},
\label{eq:cf_seq}
\end{equation}
where $a_{jk}=\langle j|z|k\rangle |\tilde E(\omega_{jk})|$ is the product of the dipole matrix elements involving the ground $i$, intermediate $n$ and final states $f$ with the chirp-independent spectral amplitude of the pulse at the corresponding transition frequencies $\omega_{jk}=E_j-E_k$, and where  $\tilde E(\omega)=\int_{-\infty}^\infty E^+(t) e^{i \omega t} \mathrm{d}t$ results from a Fourier transform of $E^+(t)$ [the \emph{c.c.}\ part in \autoref{eq:chirped_gaussian}]. The exponential in \autoref{eq:cf_seq} corresponds to the spectral phase of the field.
The good agreement between the ionization probabilities resulting from this model [shown in \autoref{fig:fig3}(b)], 
and the ab initio ones [\autoref{fig:fig3}(a)] validates the use of the sequential approximation to map the wave packet generated by the chirped pulse. 
More interestingly, \autoref{eq:cf_seq} demonstrates the close relation between the current approach and conventional pump-probe setups \cite{Feist2011,Palacios2014}. In such schemes, the two transitions
are driven by two different pulses separated by a time delay $\Delta t$, leading to the analog expression for the ionization amplitudes, $c_f^{\mathrm{PP}} \propto \sum_n a_{fn}^{(2)}a_{ni}^{(1)} e^{-i E_n \Delta t}$, but with an important difference: For the single chirped pulse, the relative phase depends quadratically on the intermediate state energy $E_n$, while it does linearly in a pump-probe scheme. However, if the transition amplitude to intermediate states is peaked around an average value $\bar E_n$ (as in the present case, cf.~\autoref{fig:fig1}c), we can bridge this difference and make the analogy even more apparent. Expanding the energy of the intermediate states around this value, $E_n=\bar E_n + \delta_n$, one obtains
\begin{equation}
  c_f \propto \sum_n a_{fn}a_{ni} e^{-i \delta_n \Delta t_{\mathrm{e}} - i\eta T_0^2 \delta_n^2} \simeq \sum_n a_{fn}a_{ni} e^{-i \delta_n \Delta t_{\mathrm{e}}},
\end{equation}
where $\Delta t_{\mathrm{e}} = (\bar\omega_{ni} - \bar\omega_{fn}) T_0^2 \eta$ corresponds to an effective time delay, defining $\bar{\omega}_{ni}=\bar{E}_n-E_i$ and  $\bar{\omega}_{fn}=E_f-\bar{E}_n$, and the quadratic term can be neglected for sufficiently small $\delta_n$. 
It can be easily shown that for large enough $\eta$ the effective time delay matches the difference between the times when the instantaneous frequency $\omega(\eta,t)$ is resonant with the average transition energies $\bar\omega_{ni}$ and $\bar\omega_{fn}$. 
In summary, these expressions demonstrate that, within the validity of the sequential approximation, the chirped pulse acts like a conventional pump-probe setup, but with an effective time delay given by an average energy difference of the transition of interest.

In conclusion, we have shown that two-photon ionization of molecules can be manipulated by using frequency-chirped femtosecond pulses, leading to modulations of the ionization probability of more than an order of magnitude.
We have also shown that chirped pulses can be used to probe the ultrafast molecular dynamics triggered in the excited molecule by just varying the frequency chirp, which is equivalent to varying the time delay in the long awaited UV pump-UV probe schemes. This has been demonstrated by using chirped pulses with the same energy spectrum and a quadratic spectral phase, which as shown in previous works \cite{Chang2005,Scrinzi2006,Hofstetter2011,Bostedt2016}, can be easily reproduced in the lab. In this scenario, the energy distribution of the wave packet created by one-photon absorption does not vary with the chirp parameter, while the spatial distribution does. This can be retrieved from its direct mapping into the energy distribution of the charged fragments after Coulomb explosion, and is shown to be formally analogous to a conventional pump-probe scheme. Although applied to H$_2^+$ in the present work, the method should also be suitable to probe wave packet dynamics in excited states of more complex molecules. It will not only be easier to implement than UV pump-IR probe methods, where two different pulses must be synchronized, but it will also avoid the significant distortion introduced by the IR probing.

\section{Acknowledgments}
We acknowledge support from the European Research Council under ERC grants no.\ 290853 XCHEM and 290981 PLASMONANOQUANTA, European COST Action CM1204 XLIC, the Ministerio de Econom\'ia y Competitividad projects FIS2013-42002-R, FIS2016-77889-R and the ``María de Maeztu'' programme for Units of Excellence in R\&D (MDM-2014-0377), European grants MC-ITN CORINF and MC-RG ATTOTREND 268284. AP acknowledges a Ram\'on y Cajal contract from the Ministerio de Econom\'ia y Competitividad. The project has been supported with computer time allocated in Mare Nostrum BSC-RES and in the Centro de Computaci\'on Cient\'ifica UAM.

\bibliography{bibJelovina}
\end{document}